\documentclass[conference,letterpaper]{IEEEtran}
\IEEEoverridecommandlockouts

\usepackage{cite}
\usepackage{algpseudocode}
\usepackage{graphicx}
\usepackage{amsmath}
\usepackage{amsfonts}
\usepackage{epstopdf}
\usepackage{xcolor}
\usepackage{enumerate}
\usepackage{multirow}
\usepackage{transparent}
\usepackage{subfigure}
\usepackage{float}
\usepackage{amssymb}
\usepackage[]{algorithm2e}
\usepackage[letterpaper, top=0.71in, bottom=1.08in, left=0.63in, right=0.64in]{geometry}
\usepackage{setspace}
\def\BibTeX{{\rm B\kern-.05em{\sc i\kern-.025em b}\kern-.08em
		T\kern-.1667em\lower.7ex\hbox{E}\kern-.125emX}}

\DeclareMathOperator*{\E}{\mathbb{E}}

\newtheorem{theorem}{Theorem}
\newtheorem{remark}{Remark}

\begin{document}
	\begin{titlepage}
		\quad\\[1cm]
		\makeatother
		{\Huge IEEE Copyright Notice}\\[0.5cm]
		{\begin{spacing}{1.2}
				\large \copyright \ 2024 IEEE. Personal use of this material is permitted. Permission from IEEE must be obtained for all other uses, in any current or future media, including reprinting/republishing this material for advertising or promotional purposes, creating new collective works, for resale or redistribution to servers or lists, or reuse of any copyrighted component of this work in other works.
		\end{spacing}}

	\end{titlepage}

	\title{Interference Cancellation for \\OTFS-Based Over-the-Air Computation}
	\author{\IEEEauthorblockN{Xinyu Huang\IEEEauthorrefmark{1}, Henrik Hellström\IEEEauthorrefmark{1}, and Carlo Fischione\IEEEauthorrefmark{1}}
		\IEEEauthorblockA{\IEEEauthorrefmark{1}Division of Network and Systems Engineering and KTH Digital Futures, KTH Royal Institute of Technology, Sweden}}
	\maketitle
	\begin{abstract}
	This paper investigates over-the-air computation (AirComp) in the context of multiple-access time-varying multipath channels. We focus on a scenario where devices with high mobility transmit their sensing data to a fusion center (FC) for averaging. To combat the time-varying channel and Doppler effect, each device adopts orthogonal time frequency space (OTFS) modulation. After signals are received by the FC, the aggregated data undergoes demodulation and estimation within the delay-Doppler domain. We leverage the mean squared error (MSE) as a metric for the computational error of OTFS-based AirComp. We then derive the optimal transmit power at each device and signal scaling factor at FC for minimizing MSE. Notably, the performance of OTFS-based AirComp is not only affected by the noise but also by the inter-symbol interference and inter-link interference arising from the multipath channel. To counteract the interference-induced computational errors, we incorporate zero-padding (ZP)-assisted OTFS into AirComp and propose algorithms for interference cancellation. Numerical results underscore the enhanced performance of ZP-assisted OTFS-based AirComp over naive OTFS-based AirComp. 
	\end{abstract}
	
	\section{Introduction}
	The Internet of Things (IoT) leads to a pressing challenge of efficient data aggregation from dense networks of devices. The traditional data aggregation method via orthogonal channels becomes highly inefficient due to the restricted spectrum resources. Recently, over-the-air computation (AirComp) was proposed as an efficient method to overcome this issue, which leverages the signal superposition property over the multiple-access channel (MAC) to allow rapid wireless data aggregation~\cite{csahin2023survey}. Due to the unique characteristics of AirComp in integrating computation and communication, it is particularly applied in distributed machine learning \cite{yang2020federated}. 
	
	Most previous studies, e.g., \cite{yang2020federated,zhu2018mimo}, considered AirComp over time-invariant channels due to the simpler mathematical analysis. However, in the next-generation of wireless networks, there are various emerging applications in high-mobility environments, such as low-earth-orbit (LEO) satellites and high speed trains \cite{saad2019vision}, which lead to a time-varying double-selective channel. Some previous studies, e.g., \cite{cao2020optimized,fu2021uav}, have investigated AirComp over time-varying channels and estimated the aggragated signal in the time domain. As pre-processing is usually required for AirComp, \cite{cao2020optimized,fu2021uav} assumed that the transmitters always know the instantaneous channel gains and the signal estimation is performed within the channel coherence time. This requires frequent signaling with large overhead to broadcast channel state information (CSI) to transmitters. Also, for fast-varying channels, the channel coherence time can be less than the signaling time, which makes AirComp in \cite{cao2020optimized,fu2021uav} potentially impossible. Furthermore, in \cite{tegin2023federated}, the authors modulated data via orthogonal frequency-division multiplexing (OFDM). The aggregated signal is estimated on the time frequency domain. However, the high Doppler shifts can generate severe inter-carrier interference, which degrades the performance of OFDM. Also, \cite{tegin2023federated} did not consider any pre-processing or post-processing techniques for reliable AirComp.
	
	To reduce the signaling overhead and eliminate the undesired effects introduced by Doppler shift, orthogonal time frequency space (OTFS) modulation was recently proposed as a promising technique that has been shown to be successful in combating the Doppler effect and time-varying channels \cite{hadani2017orthogonal}. In OTFS, information is modulated on the delay-Doppler domain for transmission, where time-varying channels have almost constant channel gain in the delay-Doppler domain. This advantage transforms designing AirComp over time-varying channels into time-invariant channels, ensuring that the same CSI is utilized for pre-processing of all data. Moreover, channels in the delay-Doppler domain are attributed with compactness and sparsity, which effectively eliminate the inter-carrier interference. Motivated by these advantages, it is worth investigating OTFS for data modulation in AirComp. To the best of the authors' knowledge, OTFS-based AirComp has not been investigated in previous studies. 
	
	In this paper, we consider AirComp over multiple-access time-varying channels, where mobile devices simultaneously transmit the sensing data to the fusion center (FC) for averaging. For each device, the data is first modulated on the delay-Doppler domain and then transmitted over the channel. In this paper, we consider a multipath channel between each device and the FC, where different path has different delay and Doppler shifts. We note that the multipath channel was never explicitly considered in previous AirComp studies. After signals are received by the FC, the aggregatged signal is also demodulated and estimated on the  delay-Doppler domain. Our objective is minimizing the computational error, which is measured by mean squared error (MSE), by jointly optimizing the transmit power at each device and the signal scaling factor, i.e., denoising factor, at the FC. Furthermore, different  from the traditional AirComp, the MSE of OTFS-based AirComp is not only influenced by noise but also by inter-symbol interference (ISI) and inter-link interference (ILI) due to the delay and Doppler shifts among multipath channels. To mitigate the impact of interference on the MSE, we further propose using zero-padding (ZP)-assisted OTFS for data modulation, where some null symbols are placed to avoid interference. 
	
	The main contributions of this paper can be summarized as follows. First, we derive the closed-form expressions for the optimal transmit power and denoising factor that minimize MSE for OTFS-based AirComp. Second, to further improve the reliability of OTFS-based AirComp, we consider ZP-assisted OTFS for data modulation in AirComp and propose a successive interference cancellation (SIC) algorithm for the averaged data estimation. Numerical results validate the optimality of derived transmit power and denoising factor. They also show that the ZP-assisted OTFS-based AirComp achieves a much lower MSE than OTFS-based AirComp. This efficiency becomes more pronounced with higher signal-to-noise ratio (SNR) and fewer path interferences.
	\section{System Model}
	\begin{figure}[!t]
		\begin{center}
			\includegraphics[width=1\columnwidth]{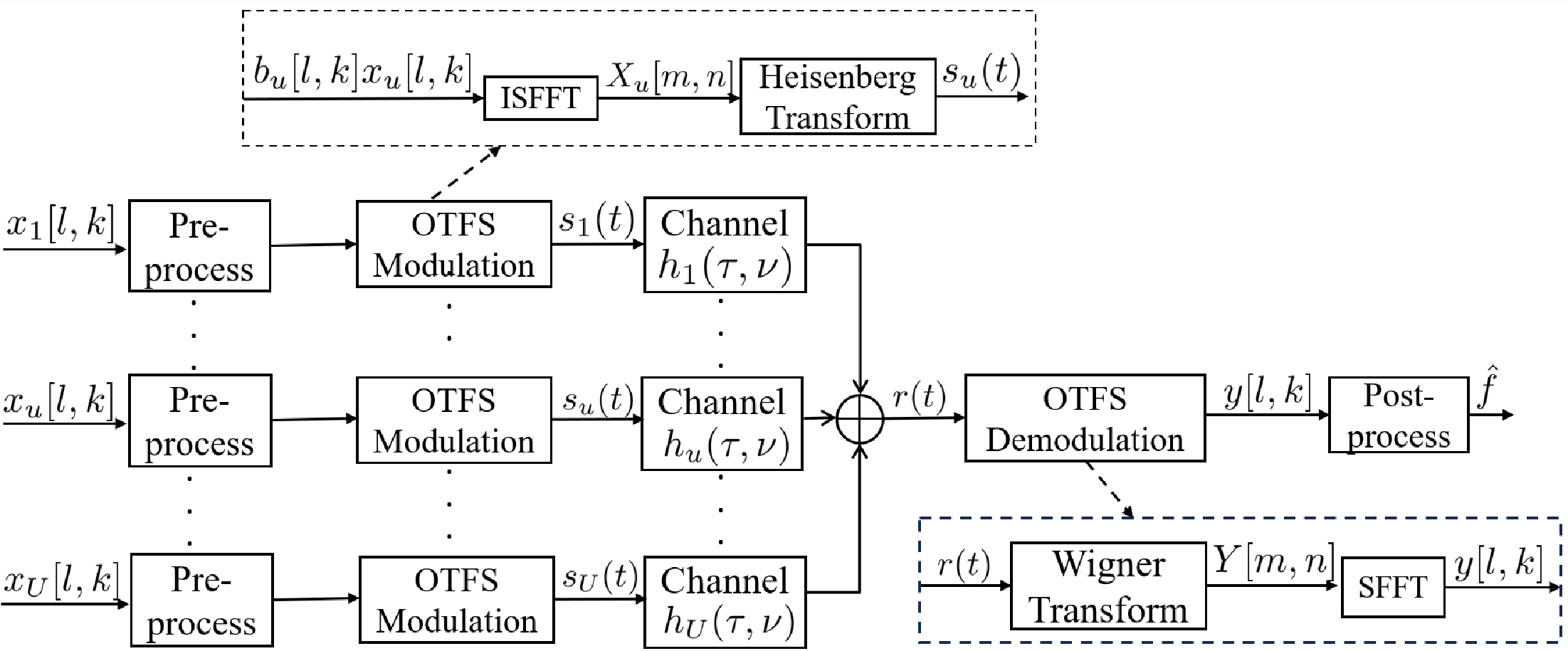}
			\caption{Illustration of the OTFS-based AirComp system.}\label{sys}\vspace{-0.5em}
		\end{center}\vspace{-4mm}
	\end{figure}

	\subsection{OTFS Modulation \& Demodulation}
	We denote $x_u[l,k]\in\mathbb{C}$ as the transmitted data, which is arranged on the $l$-th row and $k$-th column of the delay-Doppler grid for the $u$-th device. Here, $l\in\left\{0,1,..., M-1\right\}$, $k\in\left\{0,1,..., N-1\right\}$, and $u$ is any device from the set $\mathcal{U}$, where $M$ is the number of subcarriers, $N$ is the number of time slots, and $\mathcal{U}=\left\{1, ..., U\right\}$. Then, the total number of data that can be transmitted at each OTFS frame is $MN$. As in \cite{cao2020optimized}, we assume $\E\left[x_u[l,k]\right]=0$, $\E\left[x_u^2[l,k]\right]=1$, and $\E\left[x_u[l_1, k_1]x_u[l_2, k_2]\right]=0, \forall k_1\neq k_2, l_1\neq l_2$. A detailed OTFS-based AirComp system that incorporates modulation and demodulation is shown in Fig. \ref{sys}. We assume that each device and FC are equipped with a single antenna. At the transmitter side, the $u$-th device first maps $x_u[l,k]$ to $X_u[m,n]$ on the time-frequency domain using the inverse symplectic finite Fourier transform (ISFFT) as
	\begin{align}
		X_u[m,n]=\frac{1}{\sqrt{NM}}\sum_{k=0}^{N-1}\sum_{l=0}^{M-1}x_u[l,k]e^{j2\pi\left(\frac{nk}{N}-\frac{ml}{M}\right)},
	\end{align}
	where $n=0, ..., N-1$ and $m=0, ..., M-1$. Then, $X_u[m,n]$ is transformed into a continuous time waveform $s_u(t)$ by the Heisenberg transform \cite[Eq. (3)]{raviteja2018interference}.
	
	The signal $s_u(t)$ is transmitted over a time-varying channel with complex baseband channel impulse response (CIR) $h_u(\tau, \nu)$, which is given by
	\begin{align}
		h_u(\tau,\nu)=\sum_{i=1}^{R_u}h_{u,i}\delta(\tau-\tau_{u,i})\delta(\nu-\nu_{u,i}),
	\end{align}
	where $R_{u}$ is the number of propagation paths from the $u$-th device to FC, $h_{u,i}$, $\tau_{u,i}$, and $\nu_{u,i}$ represent the path gain, delay and Doppler shifts of the $i$-th path from the $u$-th device to FC, respectively, and $\delta(\cdot)$ is the Dirac delta function. We further assume that the delay and Doppler shifts are the integer multiples of $1/(M\Delta f)$ and $1/(NT)$, respectively, where $NT$ and $M\Delta f$ are the total duration and bandwidth of one OTFS frame, and we have $T\Delta f=1$ \cite{raviteja2018practical}. Accordingly, we have $\tau_{u,i}=l_{u,i}/(M\Delta f)$ and $\nu_{u,i}=k_{u,i}/(NT)$,
	where $l_{u,i}\in\mathcal{L}_u$ and $k_{u,i}\in\mathcal{K}_u$ are the indices of delay and Doppler. Here, $\mathcal{L}_u$ and $\mathcal{K}_u$ represent the sets of $l_{u,i}$ and $k_{u,i}$ among all paths between the $u$-th device and FC. Subsequently, the received signal at the FC is given by
	\begin{align}
		r(t)\!=\!\!\sum_{u=1}^{U}\!\int\!\!\int h_u(\tau, \nu)s_u(t-\tau)e^{j2\pi\nu(t-\tau)}d\tau d\nu+w(t),
	\end{align}
	where $w(t)$ is the additive white Gaussian noise (AWGN) at the FC with zero mean and variance $\sigma^2$. At the FC side, $r(t)$ first goes through a matched filter with output $Y(f,t)$, and $Y[m,n]$ is obtained by sampling $Y(f,t)$ at $f=m\Delta f$ and $t=nT$ \cite[Eqs. (7), (8)]{raviteja2018interference}. Then, $y[l,k]$ is obtained by taking the SFFT on the samples $Y[m,n]$ as
	\begin{align}
		y[l,k]=\frac{1}{\sqrt{NM}}\sum_{n=0}^{N-1}\sum_{m=0}^{M-1}Y[m,n]e^{-j2\pi\left(\frac{nk}{N}-\frac{ml}{M}\right)}.
	\end{align}
	
	In this paper, we consider a scenario where the pulse shaping filters at the transmitter and receiver adopt rectangular forms. Under this condition, the input-output relationship between $x_u[l,k]$ and $y[l,k]$ is given by \cite[Eq. (30)]{raviteja2018practical}
	\begin{align}\label{y}
		y[l,k]=&\sum_{u=1}^{U}\sum_{i=1}^{R_u}h_{u,i}\alpha_{u,i}[l,k]\tilde{x}_{u,i}+w[l,k],
	\end{align}
	where $\tilde{x}_{u,i}=x_u[[l-l_{u,i}]_M, [k-k_{u,i}]_N]$ and 
	\begin{align}\label{a}
		\alpha_{u,i}[l,k]=\left\{\begin{array}{lr}
			e^{-j2\pi\frac{k}{N}}z^{k_{u,i}([l-l_{u,i}]_M)},\;\mathrm{if}\;l\leq l_{u,i},\\
			z^{k_{u,i}([l-l_{u,i}]_M)},\;\mathrm{if}\;l\geq l_{u,i},		
		\end{array}
		\right.
	\end{align}
	with $z=e^{\frac{j2\pi}{MN}}$ and $[\cdot]_N$ denotes mod-$N$ operation. 
	\subsection{OTFS-based AirComp System}
	From \eqref{y}, we observe that $y[l,k]$ is a combination of $x_u[l,k]$ that has been shifted according to the delay and Doppler shifts specific to each channel path. Thus, for the corresponding $x_u[l,k]$ to be successfully aggregated at the FC, one succefful approach is to designate one path as the principal path. The corresponding $x_u[l,k]$ is then arranged on the delay-Doppler grid in accordance with the delay and Doppler shifts of the principal path for each device. Without any loss of generality, we consider $h_{u,1}, \forall u\in\mathcal{U}$, as the principal path gain. After pre-processing, we multiply $b_u[l,k]$ to each $x_u[l,k]$, where $b_u[l,k]$ is the transmit coefficient. Accordingly, we set $b_u[l,k]=\sqrt{p_u}h_{u,1}^\dag\alpha_{u,1}^\dag[l,k]/|h_{u,1}|$, where  $p_u\geq0$ is the transmit power of the $u$-th device and $h_{u,1}^\dag$ is the complex conjugate of $h_{u,1}$. By replacing $x_u[l,k]$ in \eqref{y} with $b_u[l,k]x_u[l,k]$, we rephrase \eqref{y} as 
	\begin{align}\label{ykls}
		y[l,k]=&\sum_{u=1}^{U}\!\!\sqrt{p_u}|h_{u,1}|\tilde{x}_{u,1}\!+\!\sum_{u=1}^{U}\!\sum_{i=2}^{R_u}\!\sqrt{p_u}h_{u,i}\notag\\&\times\frac{\alpha_{u,i}[l,k]\alpha_{u,1}^\dag[l,k] h_{u,1}^\dag}{|h_{u,1}|}\tilde{x}_{u,i}+w[l,k].
	\end{align}
	
	In \eqref{ykls}, we observe that only the time-invariant path gain, delay shift, and Doppler shift of the principal path are required for pre-processing at each device, instead of instantaneous CSI. Moreover, we observe that compared to the conventional input-output relationship of the AirComp system, the OTFS-based AirComp system suffers from additional ISI and ILI from other devices, which are caused by the delay and Doppler spread of multipath channels. In this paper, our aim is to estimate the average of the transmitted values from devices, i.e., $f[l,k]=\sum_{u=1}^{U}\tilde{x}_{u,1}/U, \forall l\in\left\{1, ..., M-1\right\}, k\in\left\{1, ..., N-1\right\}$. We note that the arrangement of $x_u\left[l, k\right]$ is based on $l_{u,1}$ and $k_{u,1}$ such that the corresponding $x_u[l,k]$ can be added together. After obtaining $y[l,k]$, the FC estimates $f[l,k]$ as $\hat{f}[l,k]=y[l,k]/(U\sqrt{\eta})$, where $\eta\geq0$ is the denoising factor. We further denote $\epsilon$ as the mean-squared error (MSE) between $\hat{f}[l,k]$ and $f[l,k]$, which is given by
	\begin{align}\label{ep}
		\epsilon=\frac{1}{U^2}\E\left[\left(\frac{y[l,k]}{\sqrt{\eta}}-\sum_{u=1}^{U}\tilde{x}_{u,1}\right)^2\right],
	\end{align}
	where the expectation is taken over $x_u[l,k]$ and noise. By substituting \eqref{ykls} into \eqref{ep}, we obtain 
	\begin{align}\label{mm}
		\epsilon=\sum_{u=1}^{U}\left(\frac{\sqrt{p_u}|h_{u,1}|}{\sqrt{\eta}}-1\right)^2+\sum_{u=1}^{U}\sum_{i=2}^{R_u}\frac{p_u|h_{u,i}|^2}{\eta}+\frac{\sigma^2}{\eta},
	\end{align}
	where we omit $1/U^2$ for simplicity.
	\section{OTFS-Based AirComp Design}\label{o}
	In this section, we jointly optimize the transmit power $p_u$ and denoising factor $\eta$ to minimize the MSE defined in \eqref{mm}. Assuming each user operates under an average power budget denoted by $P$, we formulate the optimization problem, referred to as $\mathcal{P}1$, as follows
	\begin{align}\label{mp}
		\mathcal{P}1: \!\!\!\!\!&\min_{p_u\geq0, \eta\geq0} \sum_{u=1}^{U}\!\!\left(\frac{\sqrt{p_u}|h_{u,1}|}{\sqrt{\eta}}-1\right)^2\!+\!\sum_{u=1}^{U}\sum_{i=2}^{R_u}\frac{p_u|h_{u,i}|^2}{\eta}+\frac{\sigma^2}{\eta},\notag\\&
		\mathrm{s.t.}\; p_u\leq P, \forall u\in\mathcal{U}.
	\end{align}

	We first optimize $p_u$ for any given $\eta\geq0$. Accordingly, $\mathcal{P}1$ is simplified as
	\begin{align}\label{mpu}
		&\min_{p_u\geq0} \sum_{u=1}^{U}\left(\frac{\sqrt{p_u}|h_{u,1}|}{\sqrt{\eta}}-1\right)^2+\sum_{u=1}^{U}\sum_{i=2}^{R_u}\frac{p_u|h_{u,i}|^2}{\eta},\notag\\
		&\mathrm{s.t.}\; p_u\leq P, \forall u\in\mathcal{U}.
	\end{align}
	By taking the derivative of \eqref{mpu} with respect to $p_u$, we calculate the optimal $p_u$, denoted by $\hat{p}_u$, that minimizes \eqref{mpu} without accounting for the power constraint as $\hat{p}_u=|h_{u,1}|^2\eta/(\sum_{i=1}^{R_u}|h_{u,i}|^2)^2$.
	Upon incorporating the power constraint, the optimal transmit power, denoted by $p_u^*$, is given by
	\begin{align}\label{p}
		p_u^*=\min\left(P, \frac{|h_{u,1}|^2\eta}{\left(\sum_{i=1}^{R_u}|h_{u,i}|^2\right)^2}\right).
	\end{align}
	
	We then optimize $\eta$. By substituting \eqref{p} into \eqref{mp}, we obtain the following optimization problem, denoted by $\mathcal{P}2$, as 
	\begin{align}
		&\mathcal{P}2: \min_{\eta\geq0}\sum_{u=1}^{U}\left(\min\left(\frac{\sqrt{P}|h_{u,1}|}{\sqrt{\eta}}, \frac{|h_{u,1}|^2}{\sum_{i=1}^{R_u}|h_{u,1}|^2}\right)-1\right)^2\notag\\&+\sum_{u=1}^{U}\sum_{i=2}^{R_u}\min\left(\frac{P|h_{u,i}|^2}{\eta}, \left(\frac{|h_{u,1}||h_{u,i}|}{\sum_{i=1}^{R_u}|h_{u,i}|^2}\right)^2\right)+\frac{\sigma^2}{\eta}.
	\end{align}
	Without loss of generality, we assume that
	\begin{align}
		\frac{\sum_{i=1}^{R_1}\!\!|h_{1,i}|^2}{|h_{1, 1}|}\!\!\leq\!...\!\leq\!\frac{\sum_{i=1}^{R_u}|h_{u,i}|^2}{|h_{u, 1}|}\leq...\leq\frac{\sum_{i=1}^{R_U}|h_{U,i}|^2}{|h_{U, 1}|}.
	\end{align}
	To solve $\mathcal{P}2$, we divide the value of $\eta$ into $U+1$ intervals and define the $u$-th interval, $\forall u\in\left\{0, \mathcal{U}\right\}$, as
	\begin{align}\label{I}
		I_u\!\!=\!&\left\{\!\!\eta\Bigg|\!P\!\!\left(\!\!\frac{\sum_{i=1}^{R_u}\!\!|h_{u,i}|^2}{|h_{u, 1}|}\!\!\right)^2\!\!\!\!\leq\eta\!\leq\! P\!\left(\!\frac{\sum_{i=1}^{R_{u+1}}\!\!|h_{u+1,i}|^2}{|h_{u+1, 1}|}\!\right)^2\!\right\}\!,
	\end{align}
	where we define $\sum_{i=1}^{R_0}|h_{0,i}|^2/|h_{0, 1}|=0$ and $\sum_{i=1}^{R_{U+1}}|h_{U+1,i}|^2/|h_{U+1, 1}|=\infty$. Then, for $\eta\in I_u$, $\mathcal{P}2$ can be rephrased as 
	\begin{align}\label{me}
		&\min_{\eta\in I_u}\!H_u(\eta)\!=\!\!\! \sum_{j=1}^{u}\!\!\left(\!\!\frac{\sqrt{P}|h_{j,1}|}{\sqrt{\eta}}\!-\!1\!\!\right)^2\!\!+\!\!\!\sum_{j=u+1}^{U}\!\!\left(\frac{|h_{j,1}|^2}{\sum_{i=1}^{R_j}|h_{j,i}|^2}-1\!\right)^2\notag\\&+\!\!\sum_{j=1}^{u}\!\sum_{i=2}^{R_j}\!\frac{P|h_{j,i}|^2}{\eta}\!+\!\sum_{j=u+1}^{U}\sum_{i=2}^{R_j}\frac{|h_{j,1}|^2|h_{j,i}|^2}{\left(\sum_{i=1}^{R_u}|h_{u,i}|^2\right)^2}+\frac{\sigma^2}{\eta}.
	\end{align}
	We denote $\hat{\eta}_u>0$ as the optimal $\eta$ that achieves the minimum $H_u(\eta)$, where $H_u(\eta)$ first decreases on $[0, \hat{\eta}_u]$, and then increases on $[\hat{\eta}_u, \infty]$. By computing $\partial H_u(\eta)/\partial\eta=0$, we obtain $\hat{\eta}_u$ as
	\begin{align}\label{h}
		\hat{\eta}_u=\left(\frac{\sum_{j=1}^{u}\sum_{i=1}^{R_j}P|h_{j,i}|^2+\sigma^2}{\sum_{j=1}^{u}\sqrt{P}|h_{j,1}|}\right)^2.
	\end{align}
	We then denote $\eta_u^*\in I_u$ as the optimal $\eta$ that minimizes \eqref{me}. According to \eqref{I} and \eqref{h}, $\eta_u^*$ is given by
	\begin{align}\label{et}
		\!\eta_u^*\!=\!\min\!\!\left[\!P\!\!\left(\!\frac{\sum_{i=1}^{R_{u+1}}\!|h_{u+1,i}|^2}{|h_{u+1, 1}|}\!\!\right)^2\!\!, \max\!\!\left(\!\!\hat{\eta}_u, P\!\!\left(\!\!\frac{\sum_{i=1}^{R_u}\!\!|h_{u,i}|^2}{|h_{u, 1}|}\!\!\right)^2\!\right)\!\!\right]\!\!.
	\end{align}
	We further denote $\eta^*$ as the optimal $\eta$ for $\mathcal{P}2$, which can be determined by comparing the values of $H_u(\eta_u^*)$ for $\eta_u^*$ within each interval. We suppose that $\eta^*=\eta^*_{u^*}$ and obtain $u^*$ as $u^*=\arg\min_{u\in\mathcal{U}}H_u(\eta_u^*)$. Then, based on \eqref{p},  \eqref{I}, and \eqref{et}, we derive and present $p_u^*$ and $\eta^*$ in the following theorem.
	\begin{theorem}\label{t1}
		The optimal $p_u$ and $\eta$ that jointly minimize $\mathcal{P}1$ are given by
		\begin{align}\label{pu}
			p_u^*=\left\{\begin{array}{lr}
				P, \forall u\in\left\{1, ..., u^*\right\},\\
				\frac{|h_{u,1}|^2\eta^*}{\left(\sum_{i=1}^{R_u}|h_{u,i}|^2\right)^2}, \forall u\in\left\{u^*+1, ..., U\right\},
			\end{array}
			\right.
		\end{align}
		and
		\begin{align}
			\eta^*=\left(\frac{\sum_{j=1}^{u^*}\sum_{i=1}^{R_j}P|h_{j,i}|^2+\sigma^2}{\sum_{j=1}^{u^*}\sqrt{P}|h_{j,1}|}\right)^2.
		\end{align} 
	\end{theorem}
	\begin{IEEEproof}
		The proof of Theorem \ref{t1} is similar to \cite[Appendix C]{cao2020optimized}, and we omit the details here due to the page limit.
	\end{IEEEproof}
	
	\section{SIC for ZP-Assisted OTFS-Based AirComp}
	In this section, we introduce the ZP-assisted OTFS and investigate its application in AirComp. We propose a novel algorithm for the ZP-assisted OTFS-based AirComp framework, which applies the SIC
	during aggregated signal estimation at the FC. In this algorithm, we optimize the transmit power, the denoising factor, and a coefficient for interference cancellation.
	\subsection{ZP-Assisted OTFS}
	\begin{figure}[!t]
		\begin{center}
			\includegraphics[width=1\columnwidth]{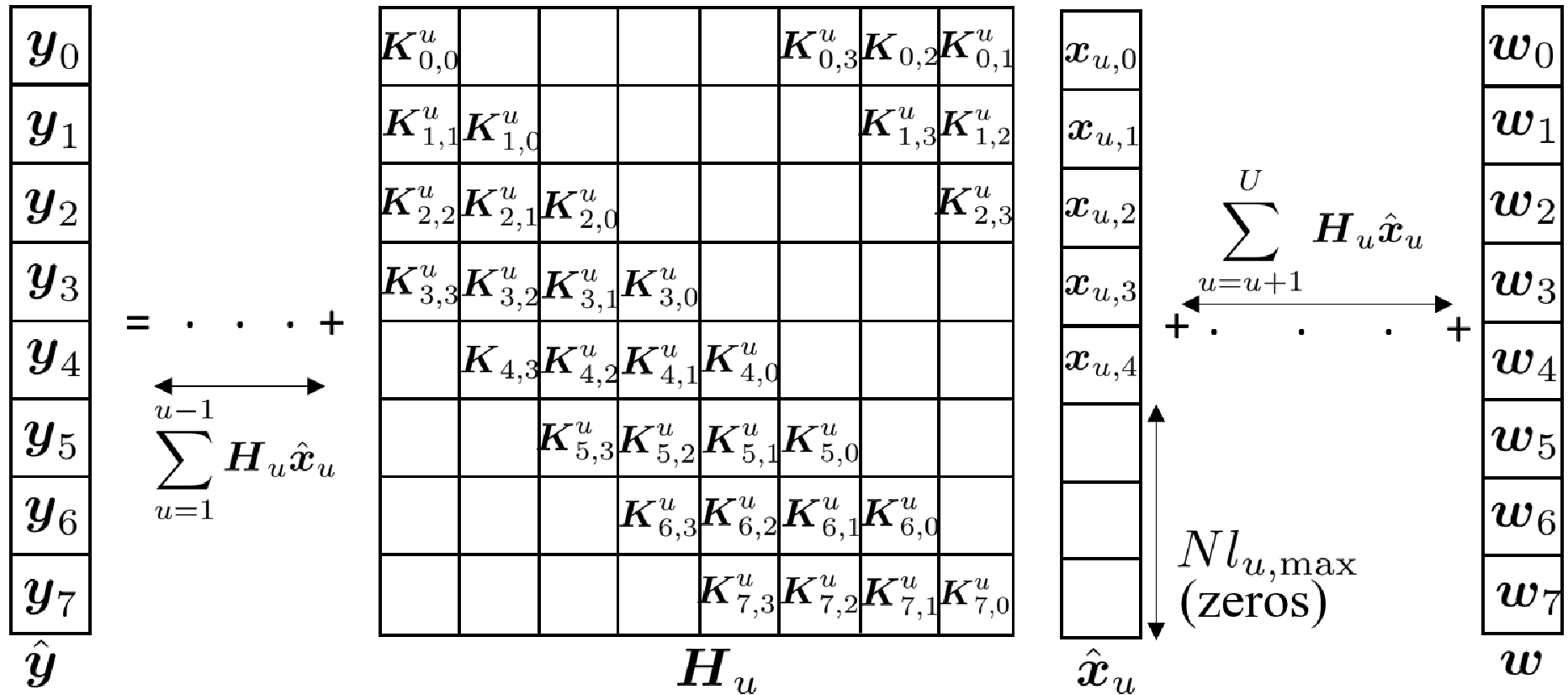}
			\caption{Delay-Doppler domain input-output relation for $N=M=8$, $R_u=4$, $l_u=[0,1,2,3]$, and $k_u=[0,1,2,3], \forall u\in\mathcal{U}$.}\label{mf}\vspace{-0.5em}
		\end{center}\vspace{-4mm}
	\end{figure}
	We denote $\boldsymbol{X}_u\in\mathbb{C}^{M\times N}$ and $\boldsymbol{Y}\in\mathbb{C}^{M\times N}$ as the two-dimensional (2D) transmitted and received data matrices in the delay-Doppler grid, respectively. In \cite{thaj2020low}, the authors proposed ZP-assisted OTFS, where the data in the last $l_{u, \mathrm{max}}$ rows of $\boldsymbol{X}_u$ are set to zero to avoid inter-block interference in the time domain. Here, $l_{u, \mathrm{max}}$ is the maximum channel delay spread index between the $u$-th device and FC. We further denote $\boldsymbol{x}_{u,m}\in\mathbb{C}^{N\times 1}$ and $\boldsymbol{y}_m\in\mathbb{C}^{N\times 1}$ as the column vectors that contain the data in the $m$-th row of $\boldsymbol{X}_u$ and $\boldsymbol{Y}$, respectively, i.e., $\boldsymbol{x}_{u,m}=\left[\boldsymbol{X}_u[m,0], \boldsymbol{X}_u[m,1], ..., \boldsymbol{X}_u[m, N-1]\right]^\mathrm{T}$ and $\boldsymbol{y}_m=\left[\boldsymbol{Y}[m,0], \boldsymbol{Y}[m,1], ..., \boldsymbol{Y}[m, N-1]\right]^\mathrm{T}$. We denote $\hat{\boldsymbol{x}}_u\in\mathbb{C}^{NM\times 1}$ and $\hat{\boldsymbol{y}}\in\mathbb{C}^{NM\times 1}$ as the vectors of the transmitted data at the $u$-th device and the received data at the FC, respectively, where $\hat{\boldsymbol{y}}=\left[\boldsymbol{y}_0^\mathrm{T}, \boldsymbol{y}_1^\mathrm{T}, ..., \boldsymbol{y}_{M-1}^\mathrm{T}\right]^\mathrm{T}$ and $\hat{\boldsymbol{x}}_u=\left[\boldsymbol{x}_{u,0}^\mathrm{T}, \boldsymbol{x}_{u,1}^\mathrm{T}, ..., \boldsymbol{x}_{u,M-1}^\mathrm{T}\right]^\mathrm{T}$. According to \cite[Eq. (30)]{thaj2020low}, the relationship between $\boldsymbol{\hat{x}}_u$ and $\hat{\boldsymbol{y}}$ is $\hat{\boldsymbol{y}}=\sum_{u=1}^{U}\boldsymbol{H}_u\hat{\boldsymbol{x}}_u+\boldsymbol{w}$, where $\boldsymbol{H}_u\in\mathbb{C}^{MN\times MN}$ is the channel matrix between the $u$-th device and FC in the delay-Doppler domain and $\boldsymbol{w}$ is the AWGN vector with zero mean and variance $\sigma^2$ for each element. In Fig. \ref{mf}, we show an configuration example, where $\boldsymbol{K}^u_{m, l}\in\mathbb{C}^{N\times N}$ within $\boldsymbol{H}_u$ for $l\in\mathcal{L}_u$. According to \cite{thaj2020low}, $\boldsymbol{K}_{m,l}^u$ is given by
	 $\boldsymbol{K}_{m,l}^u=\mathrm{circ}[\boldsymbol{\nu}_{m,l}^u(0), ..., \boldsymbol{\nu}_{m,l}^u(N-1)]$, where $\mathrm{circ}[\cdot]$ is the circulant matrix  and
	\begin{align}\label{bn}
		\boldsymbol{\nu}_{m,l}^u(\kappa)\!=\!\left\{\begin{array}{lr}
			\!\!h_{u,i}z^{k(m-l)},\mathrm{if}\;l\!=\!l_{u,i}, k\!=\!k_{u,i}, \mathrm{and}\;\kappa=[k_{u,i}]_N,\\
			\!\!0,\mathrm{otherwise}.
		\end{array}
		\right.
	\end{align}
	From Fig. \ref{mf}, it is evident that the configuration of $\boldsymbol{H}_u$ is influenced by both the number of channel paths and the respective delay and Doppler shifts within each path. To ensure an accurate aggregation of $\hat{\boldsymbol{x}}_u$ and effective SIC, we assume that $h_u(\tau,\nu), \forall u\in\mathcal{U}$, has an identical number of paths, as well as consistent delay and Doppler shifts in each of those paths, i.e.,  $R_1=...=R_U=R$, $l_{1,i}=...=l_{U,i}=l_i$, and $k_{1,i}=...=k_{U,i}=k_i$.
	\subsection{Algorithm for Interference Cancellation}\label{ef}
	As the last $Nl_{\mathrm{max}}$ elements in $\hat{\boldsymbol{x}}_u$ are always set as zero as shown in Fig. \ref{mf}, $\sum_{u=1}^{U}\boldsymbol{x}_{u,0}$ and $\sum_{u=1}^{U}\boldsymbol{x}_{u, M-l_\mathrm{max}-1}$ can always be estimated without any interference. Moreover, as the estimation of $\sum_{u=1}^{U}\boldsymbol{x}_{u,2}$ in Fig. \ref{mf} experiences the most interference, its estimation is deferred to the last position. Therefore, a prioritized sequence for estimating $\sum_{u=1}^{U}\boldsymbol{x}_{u,m}, \forall m\in\left\{0, ..., M-l_\mathrm{max}-1\right\}$, is $\left[\boldsymbol{x}_{u,0},\! \boldsymbol{x}_{u, 1},\!...,\! \boldsymbol{x}_{u,m^*}, \!x_{u, M\!-l_\mathrm{max}-1}, x_{u, M-l_\mathrm{max}-2}, ..., \boldsymbol{x}_{u,m^*+1}\right]$, where $m^*$ is determined by delays among all paths. The process for determining $m^*$ is described in following remark.
	\begin{remark}\label{w}
		We apply integer $\theta_m$ to measure the level of impact of the interference on the estimation of $\sum_{u=1}^{U}\boldsymbol{x}_{u,m}$.
		For the estimation of $\sum_{u=1}^{U}\boldsymbol{x}_{u,m_1}$, the interference is either within the interval  $[\boldsymbol{x}_{u,0}, \boldsymbol{x}_{u,m_1-1}]$ or within $[\boldsymbol{x}_{u,m_1+1}, \boldsymbol{x}_{u,M-l_\mathrm{max}-1}]$. When the interference is within $[\boldsymbol{x}_{u,0}, \boldsymbol{x}_{u,m_1-1}]$, we set $\theta_{m_1}=\theta_{m_1}^+$, and assume that there are $|\mathcal{M}^+|$ interferences $\boldsymbol{x}_{u,m}$, $\forall m\in\mathcal{M}^+$, where $\mathcal{M}^+$ is a set containing the index $m$ of all interference terms within $[\boldsymbol{x}_{u,0}, \boldsymbol{x}_{u,m_1-1}]$. When the interference is within $[\boldsymbol{x}_{u,m_1+1}, \boldsymbol{x}_{u,M-l_\mathrm{max}-1}]$, we set $\theta_{m_1}=\theta_{m_1}^-$, and assume that there are $|\mathcal{M}^-|$ interferences $\boldsymbol{x}_{u,m}$, $\forall m\in\mathcal{M}^-$. Then, we calculate $\theta_{m_1}^+$ and $\theta_{m_1}^-$ as $\theta_{m_1}^+=\sum_{m\in\mathcal{M}^+}\theta_m^++|\mathcal{M}^+|$ and $\theta_{m_1}^-=\sum_{m\in\mathcal{M}^-}\theta_m^-+|\mathcal{M}^-|$. When there is no interference for estimation, we set $\theta_{m_1}^+=\theta_{m_1}^-=0$. After obtaining $\theta_{m_1}^+$ and $\theta_{m_1}^-$, $m^*$ is determined as the minimum $m_1$ when $\theta_{m_1}^+\leq\theta_{m_1}^-$.
	\end{remark}
	
	We consider the scenario that $l_i$ on each path is different, i.e., $l_i\neq l_j, \forall i\neq j, 1\leq i,j\leq R$.\footnote{The scenario that at least two paths share the same delay will be investigated in the extended version.} Then, only one element in each row of $\boldsymbol{K}_{m,l}^u$ is non-zero. We first consider the scenario with \textit{no interference} during the estimation of $\sum_{u=1}^{U}\boldsymbol{x}_{u,m}$. To illustrate this, we use the estimation of $\sum_{u=1}^{U}\boldsymbol{x}_{u,0}$ from Fig. \ref{mf} as an example. We define $\gamma_m\in\left\{mN, ..., (m+1)N-1\right\}$ and denote $y_{\gamma_m}$, $x_{u,\gamma_m}$, and $w_{\gamma_m}$ as the $\gamma_m$th element in $\hat{\boldsymbol{y}}$, $\hat{\boldsymbol{x}}_u$, and $\boldsymbol{w}$, respectively.
	As only $\boldsymbol{\nu}_{0,0}^u(0)=h_{u,1}$ in $\boldsymbol{K}_{0,0}^u$ is non-zero, the relationship between $y_{\gamma_0}$ and $x_{u, \gamma_0}$ is given by \begin{align}\label{yg}
		y_{\gamma_0}=\sum_{u=1}^{U}\sqrt{p_{u,0}}|h_{u,1}|x_{u,\gamma_0}+w_{\gamma_0},
	\end{align}
	where $\sqrt{p_{u, m}}$ is the transmit power for $x_{u,\gamma_m}$. We denote $\hat{f}_{m_1}=y_{\gamma_{m_2}}/\sqrt{\eta_{m_2}}$ as the estimated value of $\sum_{u=1}^{U}x_{u,\gamma_{m_1}}$ and  $\epsilon_{m_1}$ as the MSE associated with estimating $\sum_{u=1}^{U}x_{u, \gamma_{m_1}}$, where $\eta_{m_2}$ is the denoising factor for $y_{\gamma_{m_2}}$. We note that \eqref{yg} has a same format as that of \cite[Eq. (5)]{cao2020optimized}, which suggests that the optimization procedures of $p_{u,0}$ and $\eta_0$ can follow \cite{cao2020optimized}. Thus, $p_{u,0}^*$ and $\eta_0^*$ are given by \cite[Eqs. (24), (25)]{cao2020optimized}
	\begin{align}\label{pug}
		p^*_{u, 0}=\left\{\begin{array}{lr}
			P,\;\forall u\in\left\{1, ..., u_0^*\right\},\\
			\frac{\eta^*_0}{|h_{u,1}|^2}, \forall u\in\left\{u_0^*+1, ..., U\right\},
		\end{array}
		\right.
	\end{align}
	and
	\begin{align}\label{e}
		\eta^*_0=\left(\frac{\sigma^2+\sum_{j=1}^{u^*_0}P|h_{j,1}|^2}{\sum_{j=1}^{u^*_0}\sqrt{P}|h_{j,1}|}\right)^2,
	\end{align}
	respectively, with $u_0^*=\min_{u}\left(\frac{\sigma^2+\sum_{j=1}^{u}P|h_{j,1}|^2}{\sum_{j=1}^{u}\sqrt{P}|h_{j,1}|}\right)^2$.
	
	Next, we consider the scenario where multiple interferences impact the estimation of $\sum_{u=1}^{U}\boldsymbol{x}_{u,m}$. For illustrative purposes, we focus on the estimation of $\sum_{u=1}^{U}\boldsymbol{x}_{u,1}$ as an example. As depicted in Fig. \ref{mf}, $x_{u,\gamma_0}$ acts as an interference during the estimation of $\sum_{u=1}^{U}x_{u,\gamma_1}$ when examining the relationship between $x_{u,\gamma_1}$ and $y_{\gamma_1}$. Given that we have estimated $\sum_{u=1}^{U}x_{u,\gamma_0}$, it can be subtracted from $y_{\gamma_1}$, leading to
	\begin{align}\label{yg1}
		y_{\gamma_1}=&\sum_{u=1}^{U}\sqrt{p_{u, 1}}|h_{u,1}|x_{u,\gamma_1}+\sum_{u=1}^{U}\frac{\sqrt{p_{u,0}^*}h_{u,2}h_{u,1}^\dag}{|h_{u,1}|}x_{u, \gamma_0}\notag\\&+w_{\gamma_1}-\zeta_{1,0}\hat{f}_0,
	\end{align}
	where $\zeta_{m,j}>0$ is a coefficient for $\hat{f}_j$ subtracted from the signal $y_{\gamma_m}$. By substituting \eqref{yg1} and $\hat{f}_0$ into \eqref{ep}, we formulate the optimization problem that minimizes the MSE between $\hat{f}_1$ and $\sum_{u=1}^{U}x_{u,\gamma_1}/U$ as
	\begin{align}\label{eps}
		\mathcal{P}3: &\min_{p_{u,1}\geq0, \eta_1\geq0, \zeta_{1,0}\geq0}	\epsilon_1=\!\sum_{u=1}^{U}\!\left(\frac{p_{u, 1}|h_{u,1}|}{\sqrt{\eta_1}}-1\right)^2\notag\\&+\frac{\E[G(\zeta_{1,0})^2]}{\eta_1}, \;\;\mathrm{s.t.}\;p_{u,1}\leq P, \forall u\in\mathcal{U},
	\end{align}
	where 
	\begin{align}
		G(\zeta_{1,0})=&\sum_{u=1}^{U}\left(\frac{h_{u,2}\sqrt{p_{u, 0}^*}h^\dag_{u,1}}{|h_{u,1}|}-\frac{\zeta_{1,0}\sqrt{p_{u, 0}^*}|h_{u,1}|}{\sqrt{\eta_0^*}}\right)x_{u, \gamma_0}\notag\\&+w_{\gamma_1}-\frac{\zeta_{1,0}w_{\gamma_0}}{\sqrt{\eta_0^*}}.
	\end{align}
	In the following theorem, we derive and present the optimal $p_{u,1}$, $\eta_1$, and $\zeta_{1,0}$ for $\mathcal{P}3$.
	\begin{theorem}\label{t2}
		The optimal $p_{u, 1}$, $\eta_1$, and $\zeta_{1,0}$ that jointly minimize $\mathcal{P}3$ are given by
		\begin{align}\label{pug1}
			p_{u,1}^*=\left\{\begin{array}{lr}
				P,\;\forall k\in\left\{1, ..., u_1^*\right\},\\
				\frac{\eta^*_1}{|h_{u,1}|^2}, \forall u\in\left\{u_1^*+1, ..., U\right\},
			\end{array}
			\right.
		\end{align}
	\begin{align}\label{egl}
		\eta_1^*=\left(\frac{\min\E[G(\zeta_{1,0})^2]+\sum_{j=1}^{u^*_1}P|h_{j,1}|^2}{\sum_{j=1}^{u^*_1}\sqrt{P}|h_{u,1}|}\right)^2,
	\end{align}
		and 
		\begin{align}\label{d}
			\zeta_{1,0}^*=\frac{\sqrt{\eta_0^*}\sum_{u=1}^{U}|h_{u,2}||h_{u,1}|p_{u,0}^*}{\sum_{u=1}^{U}p_{u,0}^*|h_{u,1}|^2+\sigma^2},
		\end{align}
		where
		\begin{align}
			\min\E[G(\zeta_{1,0})^2]=&\sum_{u=1}^{U}|h_{u,2}|^2p_{u,0}^*+\sigma^2\notag\\&-\frac{(\sum_{u=1}^{U}|h_{u,2}||h_{u,1}|p_{u,0}^*)^2}{\sum_{u=1}^{U}p_{u,0}^*|h_{u,1}|^2+\sigma^2},
		\end{align}
		and
		\begin{align}
			u_1^*=\min_u\left(\frac{\min\E\left[G(\zeta_{1,0})^2\right]+\sum_{j=1}^{u}P|h_{j,1}|^2}{\sum_{j=1}^{u}\sqrt{P}|h_{j,1}|}\right)^2.
		\end{align}
	\end{theorem}
	\begin{IEEEproof}
		Please see Appendix \ref{A2}.
	\end{IEEEproof}
	
Notably, the estimation of $\sum_{u=1}^{U}x_{u,\gamma_1}$ in \eqref{yg1} is influenced only by a single interference $x_{u,\gamma_0}$. Furthermore, we consider the scenario that the detection of $\sum_{u=1}^{U}x_{u,\gamma_{m_1}}$ is affected by more than one interference, denoted by $x_{\gamma_m}, \forall m\in\mathcal{M}$, where $\mathcal{M}$ represents the set for the index $m$ of the interference. Under this scenario, the relationship between $x_{u,\gamma_{m_1}}$ and $y_{\gamma_{m_2}}$ is given by 
	\begin{align}
		&y_{\gamma_{m_2}}\!\!\!=\!\!\!\sum_{u=1}^{U}\!\sqrt{p_{u, m_1}}|\nu_{m_2,l}^u|x_{u,\gamma_{m_1}}\!\!+\!\!\!\sum_{m\in\mathcal{M}}\sum_{u=1}^{U}\!\sqrt{p_{u,m}^*}\nu_{m_2, m_1-m+l}^u\notag\\&\times\frac{(\nu_{m_2,l}^u)^\dag x_{u,\gamma_{m}}}{|\nu_{m_2,l}^u|}+w_{\gamma_{m_2}}-\sum_{m\in\mathcal{M}}\zeta_{m_2, m}\hat{f}_{m},
	\end{align}
	where $\nu_{m_2,m_1-m+l}^u$ represents the non-zero element in $\boldsymbol{K}_{m_2,m_1-m+l}^u$ that corresponds to $x_{u,\gamma_m}$. Here, $y_{\gamma_{m_2}}$ is chosen to result in a minimum $\theta_{m_1}$ for $x_{u,\gamma_{m_1}}$. We then formulate the optimization problem based on the MSE between $\hat{f}_{m_1}$ and $\sum_{u=1}^{U}x_{u,\gamma_{m_1}}/U$ as
	\begin{align}\label{mpur}
		&\min_{p_{u,m_1}\geq0, \eta_{m_2}\geq0,\zeta_{m_2, m}\geq0, \forall m\in\mathcal{M}}\epsilon_{m_1}\!\!=\!\!\sum_{u=1}^{U}\!\!\left(\!\frac{\sqrt{p_{u, m_1}}|\nu_{m_2, l}^u|}{\sqrt{\eta_{m_2}}}\!-\!1\!\right)^2\notag\\&+\frac{\sum_{m\in\mathcal{M}}\E\left[G(\zeta_{m_2, m})^2\right]}{\eta_{m_2}},\;\mathrm{s.t.}\;p_{u, m_1}\leq P, \forall u\in\mathcal{U},
	\end{align}
	where $E\left[G(\zeta_{m_2, m})^2\right]$ is the quadratic function relating to $\zeta_{m_2, m}$. Since \eqref{mpur} has a same format as that of \eqref{eps}, optimal values $p_{u,m_1}^*$, $\eta_{m_2}^*$, and $\zeta_{m_2, m}^*$ can be derived accordingly, as shown in Appendix \ref{A2}.
	\section{Numerical Results}
	In this section, we evaluate the MSE performance of both OTFS-based AirComp and  ZP-assisted OTFS-based AirComp. The analysis focuses on the average MSE over a large number of channel realizations. Specifically, we set $M=32$, $N=16$, $U=20$, $R_u=4$, unless otherwise stated. The path gains are randomly generated based on a uniform power delay profile, and the delay and Doppler indices are randomly generated within the range of $[0, l_{\mathrm{max}}]$ and $[-k_\mathrm{max}, k_\mathrm{max}]$, where $l_{\mathrm{max}}=10$ and $k_\mathrm{max}=5$ \cite{li2021cross}. During the simulation, we always assume that $l_{u,1}\leq l_{u,2}\leq...\leq l_{u,R_u}$. 
	\begin{figure}[!t]
		\begin{center}
			\includegraphics[width=1\columnwidth]{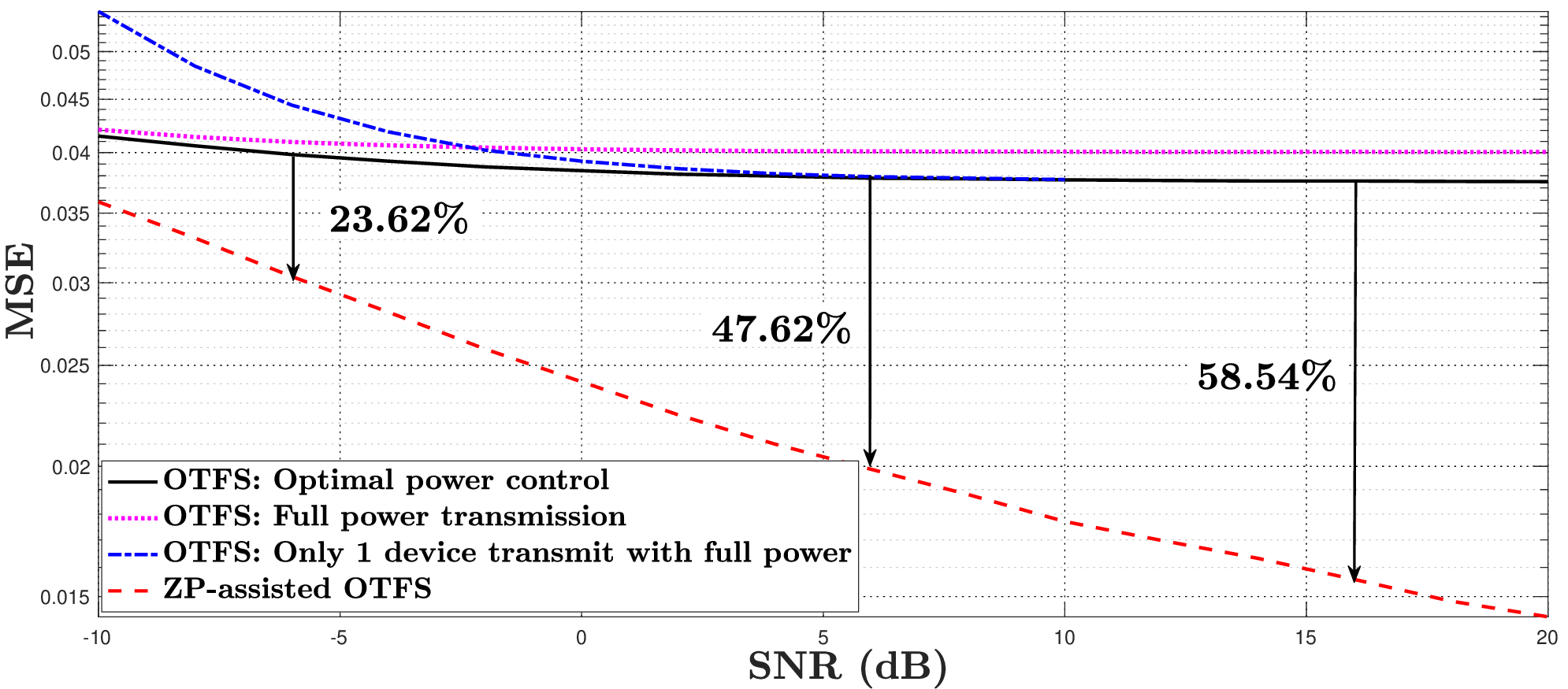}
			\caption{MSE versus SNR for OTFS-based AirComp and ZP-assisted OTFS-based AirComp, where different transmission power policies are applied.}\label{snr}\vspace{-0.5em}
		\end{center}\vspace{-4mm}
	\end{figure}
	
	In Fig. \ref{snr}, we plot MSE versus SNR for OTFS-based AirComp and ZP-assisted OTFS-based AirComp, where SNR is defined as $P/\sigma^2$. First, we apply different transmission power policies for OTFS-based AirComp. Our observations reveal that assigning optimal transmit power according to Theorem \ref{t1} results in a lower MSE compared to allotting full transmit power to all devices or to a singular device. This supports the optimal efficiency of $p_u^*$ and $\eta^*$ in minimizing MSE. Second, we observe that an MSE plateaus of OTFS-based AirComp for larger SNR, which reveals the fact that interference plays the pivotal role in degrading the performance of computation. We further observe that MSE for ZP-assisted OTFS-based AirComp is much lower than that of OTFS-based AirComp, which proves the superiority of ZP-assisted OTFS in mitigating the impact of interference. We also show that the superiority becomes more obvious as SNR increases. 
	
	\begin{figure}[!t]
		\begin{center}
			\includegraphics[width=1\columnwidth]{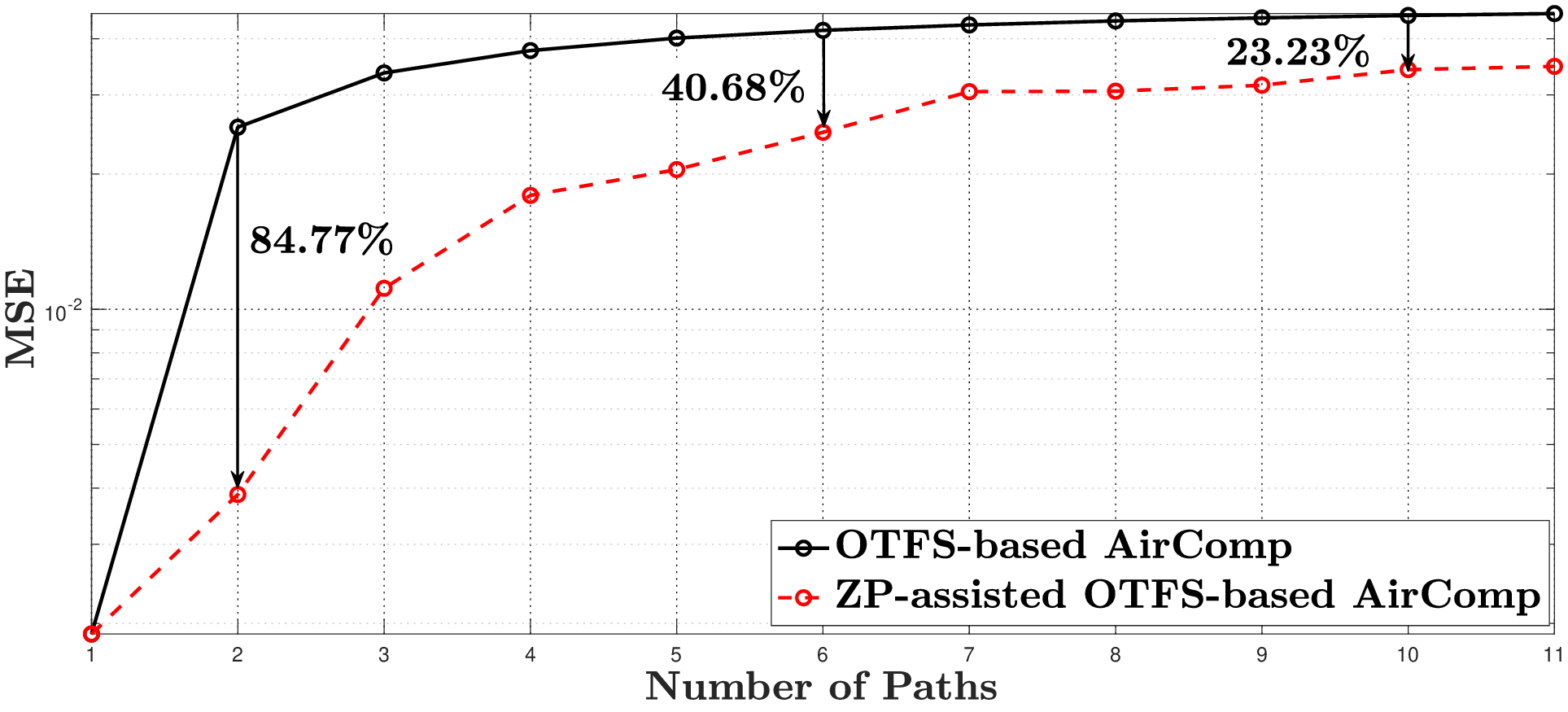}
			\caption{MSE versus number of paths for OTFS-based AirComp and ZP-assisted OTFS-based AirComp, where $\mathrm{SNR}=10\;\mathrm{dB}$.}\label{path}\vspace{-0.5em}
		\end{center}\vspace{-4mm}
	\end{figure}
	
	In Fig. \ref{path}, we depict the MSE versus the number of channel paths for both OTFS-based AirComp and ZP-assisted OTFS-based AirComp. First, our observations indicate that the MSEs for both systems increase with the increase in the number of paths. This is due to the fact that a larger number of paths introduces more interferences. Second, the gap between MSEs for both systems decreases as the number of paths increases, which demonstrates that the algorithm in Section \ref{ef} is more proficient in mitigating interference impacts when the number of channel paths is small. 
	
	\section{Conclusion}
	In this paper, we investigated OTFS-based AirComp over double-selective time-varying channels, where devices modulate data in the delay-Doppler domain before transmitting it to the FC. We derived the optimal transmit power and denoising factor that achieve the minimum MSE for the estimation of the averaged data at FC. Due to the existence of interference in OTFS-based AirComp, we further applied ZP-assisted OTFS at each device and proposed an algorithm that can effectively mitigate the interference. Numerical results showed that the performance of OTFS-based AirComp is mainly impacted by interference and that applying ZP-assisted OTFS can achieve a much lower MSE. In future research, we will integrate channel estimation into OTFS-based AirComp.
	
\appendices
	\section{Proof of Theorem \ref{t2}}\label{A2}
	We first derive $\zeta_{1,0}^*$. From \eqref{eps}, we can see that only $\E[G(\zeta_{1,0})^2]$ relates to $\zeta_{1,0}$. Therefore, we need to find $\zeta_{1,0}^*$ that minimizes $\E[G(\zeta_{1,0})^2]$. We note that the expectation of $G(\zeta_{1,0})$ is taken over $x_{u,\gamma_0}$, $w_{\gamma_1}$, and $w_{\gamma_0}$. As $x_{u,\gamma_0}$, $w_{\gamma_1}$, and $w_{\gamma_0}$ are independent with each other, $\E[G(\zeta_{1,0})^2]$ is derived as
	\begin{align}\label{egd}
		&\E[G(\zeta_{1,0})^2]=\frac{\sum_{u=1}^{U}p_{u,0}^*|h_{u,1}|^2+\sigma^2}{\eta_0^*}\zeta_{1,0}^2\notag\\&\!-\!\frac{2\sum_{u=1}^{U}|h_{u,2}||h_{u,1}|p_{u,0}^*}{\sqrt{\eta_0^*}}\zeta_{1,0}\!+\!\!\sum_{u=1}^{U}|h_{u,2}|^2p_{u,0}^*\!+\!\sigma^2.
	\end{align}
	Since \eqref{egd} is a quadratic function, the optimal $\zeta_{1,0}^*$ can be obtained by calculating $\partial\E[G(\zeta_{1,0})^2]/\partial\zeta_{1,0}=0$, which leads to \eqref{d}. By substituting $\zeta_{1,0}^*$ into \eqref{egd}, we can obtain $\min\E[G(\zeta_{1,0})^2]$. After obtaining $\zeta_{1,0}^*$, the optimization problem in \eqref{eps} becomes 
	\begin{align}\label{mpug}
		\min_{p_{u,1}\geq0, \eta_1\geq0}  &\epsilon_1=\sum_{u=1}^{U}\left(\frac{p_{u,1}|h_{u,1}|}{\sqrt{\eta_1}}-1\right)^2+\frac{\min\E[G(\zeta_{1,0})^2]}{\eta_1}, \notag\\
		&\mathrm{s.t.}\;p_{u,1}\leq P.
	\end{align}
	We note that the optimization problem in \eqref{mpug} is the same as the problem for estimating $\sum_{u=1}^{U}x_{u,\gamma_0}$ in \eqref{yg}. Then, \eqref{pug1} and \eqref{egl} can be obtained following \eqref{pug} and \eqref{e}, respectively.
\section*{Acknowledgement}
The work was supported by the following projects: KAW TAIRCOMP, SSF SAICOM, and DF DEMOCRITUS.
	\bibliographystyle{IEEEtran}
	\bibliography{ref} 
\end{document}